# A Simple and Accurate Closed-Form EGN Model Formula

P. Poggiolini, Y. Jiang, A. Carena, F. Forghieri

*Abstract*— The GN model of non-linear fiber propagation has been shown to overestimate the variance of non-linearity due to the signal Gaussianity approximation, leading to maximum reach predictions for typical optical systems which may be pessimistic by about 5% to 15%, depending on fiber type and system set-up. Various models have been proposed which improve over the GN model accuracy. One of them is the EGN model, which completely removes the Gaussianity approximation from all non-linear interference (NLI) components. The EGN model is, however, substantially more complex than the GN model. Recently, we proposed a simple closed-form formula which permits to approximate the EGN model, starting from the GN. It was however limited to all-identical, equispaced channels, and did not correct single-channel NLI (also called SCI). In this follow-up contribution, we propose an improved version which both allows to address non-identical channels and corrects the SCI contribution as well. Extensive simulative testing shows the new formula to be asymptotically very accurate, for sufficiently large number of spans. This is a preliminary document which will be followed by a journal submission.

## I. INTRODUCTION

This paper is a preliminary version which is meant to report on a significant improvement in the closed-form GN model correction formula introduced in [1]. For context, background and introduction, we refer the reader to [1].

This new formula accurately approximates the EGN model, with increasing accuracy as the number of spans grows. It includes SCI (single-channel-interference) correction, which the previous version did not include. As a result, the simulative tests presented in this paper involve the whole of NLI and not just the inter-channel components (called XMCI in [1]). Thanks to the inclusion of SCI, the system maximum reach validation now clearly shows excellent accuracy too, while in [1] a small residual error was still present.

The new formula can deal with non-identical and non-equally spaced channels. In this preliminary document the validation for this more general case is not provided, though the analytical form presented supports it.

Regarding the derivation of the formula, the version presented in [1] had its full derivation available as an appendix to that paper. The derivation of the new formula presented here simply extends that derivation. It is not included in this version of this document, but will be reported in the journal paper submission.

## II. THE EGN MODEL APPROXIMATION

Throughout the paper we assume dual-polarization propagation, over realistic fibers with non-zero loss. The EGN model [1] approximation that we propose is shown in the following. Calling $G_{\mathrm{NLI}}^{\mathrm{EGN}}(f)$ the power spectral density (PSD) of NLI noise according to the EGN model [2], it is:

$$G_{\mathrm{NLI}}^{\mathrm{EGN}}(f) \approx G_{\mathrm{NLI}}^{\mathrm{GN}}(f) - G_{\mathrm{corr}}$$
**Eq. 1**

where:

$$G_{\mathrm{corr}} = \frac{40}{81} \frac{\gamma^2 P_m N_s \bar{L}_{\mathrm{eff}}^2}{R_m \pi |\beta_2| \bar{L}_s} \left( \sum_{\substack{n=1 \\ n \neq m}}^{N_{\mathrm{ch}}} \Phi_n \frac{P_n^2}{R_n |f_n - f_m|} + \Phi_m \frac{2 P_m^2}{R_m^2} \right)$$
**Eq. 2**

with the $m$-th channel being the channel under test (CUT). It can deal with non-identical and non-equally spaced channels.

If the channels are all identical and equally spaced, and the CUT is the center channel, then Eq. 2 can be re-written as:

$$G_{\mathrm{corr}} = \frac{80}{81} \Phi \frac{\gamma^2 \bar{L}_{\mathrm{eff}}^2 P_{\mathrm{ch}}^3 N_s}{R_s^2 \Delta f \pi |\beta_2| \bar{L}_s} \cdot \left[ \mathrm{HN}\left( \frac{N_{\mathrm{ch}} - 1}{2} \right) + \frac{\Delta f}{R_s} \right]$$
**Eq. 3**

and $G_{\mathrm{NLI}}^{\mathrm{GN}}(f)$ is the NLI PSD according to the (coherent) GN model ([3], Eq. 2). The term $G_{\mathrm{corr}}$ is a closed-form 'correction' which approximately corrects the GN model for the errors due to the signal Gaussianity assumption.

The meaning of the symbols is as follows:

P. Poggiolini, Y. Jiang and A. Carena are with Dipartimento di Elettronica e Telecomunicazioni, Politecnico di Torino, Corso Duca degli Abruzzi 24, 10129, Torino, Italy, e-mail: pierluigi.poggiolini@polito.it; F. Forghieri is with CISCO Photonics, Via Santa Maria Molgora 48 C, 20871 Vimercate (MB), Italy, e-mail fforghie@cisco.com. This work was supported by CISCO Systems within a sponsored research agreement (SRA) contract.

---

[1] The EGN model is based on the Manakov equation, which accounts for the non-linear effect of one polarization on the other [4]. We use its simplified version consisting of the left-hand side of Eq. (12) in [4], which disregards polarization-mode dispersion (PMD). The linear effect of PMD is no longer a factor in modern coherent systems thanks to receiver digital signal processing (DSP). As for the non-linear impact of PMD, in [4] it was assessed to be very small or negligible in typical transmission links. Though PMD may have some impact on NLI, we consider neglecting it a reasonable approximation, for the purpose of achieving manageable analytical modeling.

- $f$ : optical frequency (THz), with $f=0$ conventionally being the center frequency of the center channel
- $\alpha$ : optical field fiber loss (1/km), such that the optical *field* attenuates as $e^{-\alpha z}$; note that the optical *power* attenuates as $e^{-2\alpha z}$
- $\beta_2$ : dispersion coefficient (ps$^2$/km)
- $\gamma$ : fiber non-linearity coefficient, 1/(W km)
- $\bar{L}_s$ : average span length (km)
- $\bar{L}_{\text{eff}}$ : average span effective length (km), with span effective length defined as $L_{\text{eff}} = \left(1 - e^{-2\alpha L_s}\right)/2\alpha$
- $N_s$ : total number of spans in the link
- $N_{\text{ch}}$ : total number of channels in the system
- $P_n$ : launch power of the *n*-th channel (W)
- $P_{\text{ch}}$ : launch power per channel for systems where the channels are all identical (W)
- $R_n$ : symbol rate of the *n*-th channel (TBaud)
- $R_s$ : symbol rate for systems where the channels are all identical (TBaud)
- $f_n$ : center frequency of the *n*-th channel (THz)
- $\Delta f$ : channel spacing for systems where the channels are all equally spaced (THz)

The specified units ensure consistency if used to express the parameters in Eq. 2. In addition, $\text{HN}(N)$ is the harmonic number series, defined as:

$$\text{HN}(N) = \sum_{n=1}^{N} (1/n)$$
**Eq. 4**

Finally, $\Phi_n$, $\Phi_m$ and $\Phi$ are constant that depend on the modulation format (see [2]). The values are: 1, 17/25 and 13/21 for PM-QPSK, PM-16QAM and PM-64QAM, respectively.

Eq. 2 assumes that the symbol rate of the CUT channel is not too low. The following relation should be satisfied: $R_m \geq \left|1/\left[\pi\beta_2 N_s L_s \left(f_n - R_n/2\right)\right]\right|$, $n = m \pm 1$. At ideal Nyquist WDM, with identical channels, the constraint simplifies to $R_s \geq \sqrt{2/(\pi|\beta_2|N_s L_s)}$. In practice, single-carrier type systems never pose any problem, whereas Eq. 2 should not be used with either OFDM or multi-subcarrier channels with very low subcarrier symbol rate.

Eq. 2 assumes that the same type of fiber is used in all spans. Spans can be of different length, though: Eq. 2 uses the average span length $\bar{L}_s$ and the average span effective length $\bar{L}_{\text{eff}}$. Accuracy is quite good for links having all individual span lengths within $\bar{L}_s \pm 15\%$. Caution should be used for larger deviations.

Eq. 2 also assumes lumped amplification, exactly compensating for the loss of the preceding span.

Eq. 2 has the following further limitations.

- $G_{\text{corr}}$ is asymptotic in the number of spans. As a result, its accuracy improves as the number of spans grows. The speed of the asymptotic convergence depends on the number of channels and on fiber dispersion (see Sect. III).
- $G_{\text{corr}}$ is derived assuming ideally rectangular channel spectra. If spectra have a significantly different shape (such as sinc-shaped), some accuracy may be lost.
- $G_{\text{corr}}$ is calculated at $f=0$ and then it is assumed to be frequency-flat.

### III. VALIDATION OF $G_{\text{corr}}$

In this section, we assume that all channels are identical and equally spaced. A straightforward choice for the quantity to focus on for model validation could be:

$$P_{\text{NLI}} = \int_{-R_s/2}^{R_s/2} G_{\text{NLI}}(f)\,df$$
**Eq. 5**

It represents the total NLI noise spectrally located within the center WDM channel. However, $P_{\text{NLI}}$ depends on the signal launch power. Specifically, it is proportional to $P_{\text{ch}}^3$. If $P_{\text{NLI}}$ is normalized with respect to $P_{\text{ch}}^3$, then the resulting quantity does not change vs. the launch power and becomes a power-independent characterization of the NLI behavior of the link. We therefore decided to concentrate on the normalized quantity $\eta_{\text{NLI}}$, defined simply as:

$$\eta_{\text{NLI}} = \frac{P_{\text{NLI}}}{P_{\text{ch}}^3}.$$
**Eq. 6**

We estimated $\eta_{\text{NLI}}$, in three ways:
1. through accurate computer simulations;
2. calculating $G_{\text{NLI}}(f)$ in Eq. 5 using the EGN model formulas for the NLI PSD provided in [2];
3. approximating $G_{\text{NLI}}(f)$ in Eq. 5 using the approximate EGN model Eq. 1.

Regarding the computer simulations, the same simulation software, simulation techniques and general system set-up described in [3], Sect. V, were used. The main details are reported in the following.

The fiber simulation algorithm is based on the standard split-step integration technique. The simulated systems symbol rate was $R_s = 32$ GBaud, with raised-cosine signal PSD and roll-off 0.05. The channel spacing was 33.6 GHz. The launch-power was -3 dBm per channel. Note that the quantity $\eta_{\text{NLI}}$ is defined so as to be launch-power independent but nonetheless we redid some of the simulations at both -6 and 0 dBm to check whether any changes could be seen. We found no significant difference.

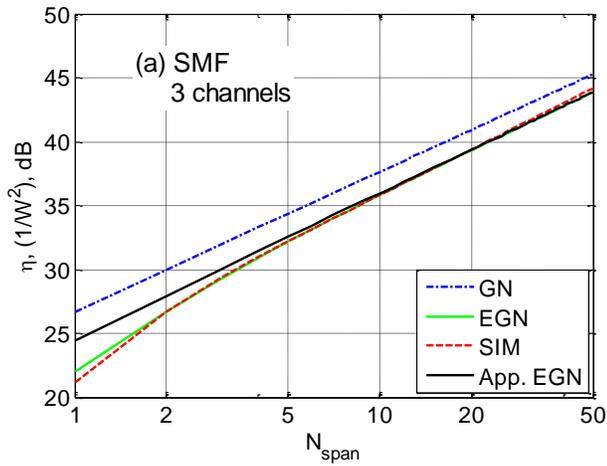
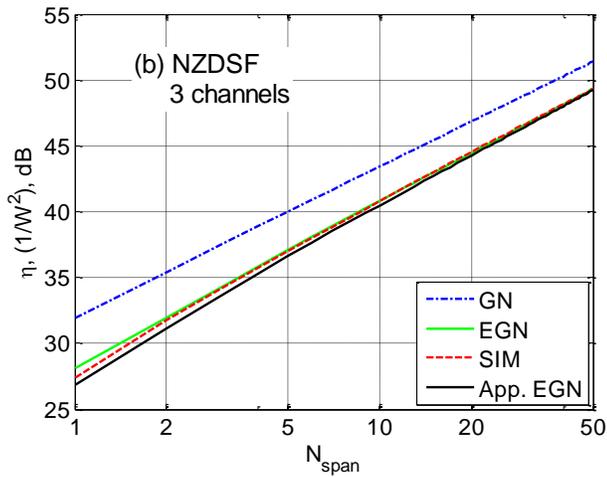
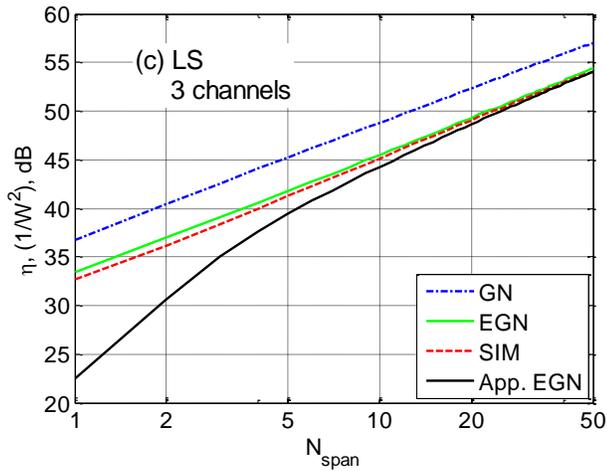

Fig. 1: Plot of the normalized NLI noise power coefficient $\eta_{NLI}$ affecting the center channel (in dB referred to $1 \cdot W^{-2}$), vs. number of spans in the link. System data: 3 PM-QPSK channels, 32 GBaud, roll-off 0.05, span length 100 km, channel spacing 33.6 GHz. The 'App. EGN' curve is generated using Eq. 1.

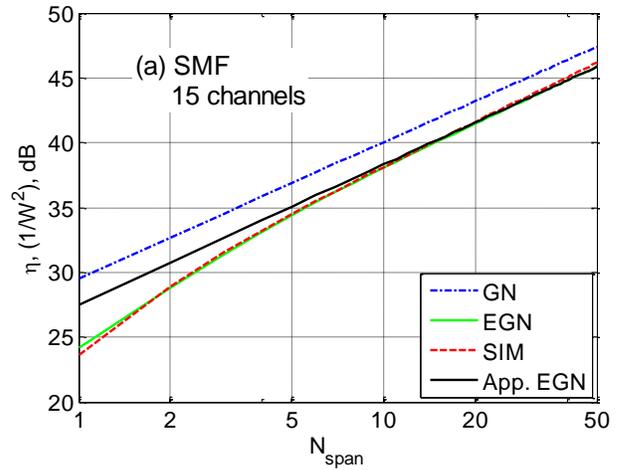
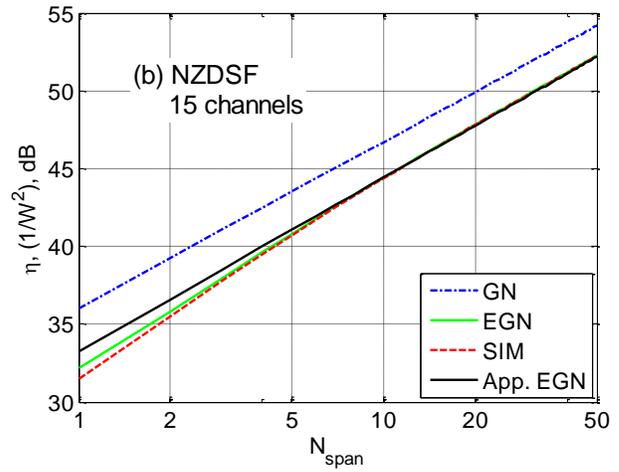
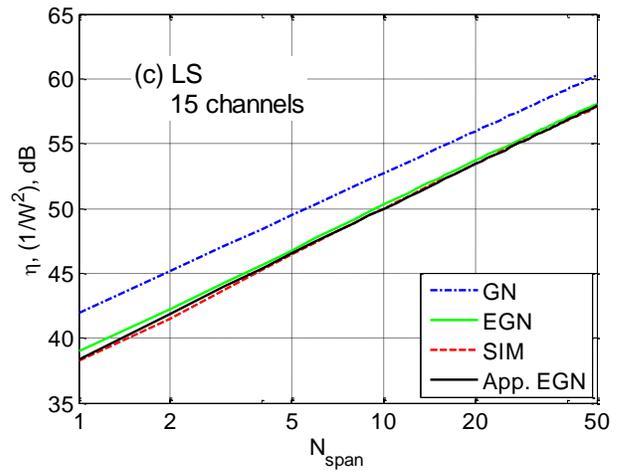

Fig. 2: Plot of the normalized NLI noise power coefficient $\eta_{NLI}$ affecting the center channel (in dB referred to $1 \cdot W^{-2}$), vs. number of spans in the link. System data: 15 PM-QPSK channels, 32 GBaud, roll-off 0.05, span length 100 km, channel spacing 33.6 GHz. The 'App. EGN' curve is generated using Eq. 1.

The tested fibers were: standard single-mode (SMF) with $D=16.7$ ps/(nm·km), $\gamma=1.3$ (W·km)$^{-1}$; non-zero dispersion-shifted fiber (NZDSF, similar to OFS's TrueWave RS), with $D=3.8$ ps/(nm·km), $\gamma=1.5$ (W·km)$^{-1}$; negative non-zero dispersion-shifted fiber (which we call "LS" because it is similar to Corning's LS fiber) with $D=-1.8$ ps/(nm·km), $\gamma=2.2$ (W·km)$^{-1}$. The span length was $L_s=100$ km and loss was $\alpha_{dB}=0.22$ dB/km for all fibers.

The Rx compensated statically for polarization rotation and applied an ideal matched filter. No dynamic equalizer was used, to avoid any possible effect of the equalizer adaptivity on NLI estimation. The simulation was completely noiseless: neither ASE noise, nor any other types of noise, such as Rx electrical noise, were present.

A first set of results is plotted in Fig. 1-Fig. 2. The quantity $\eta$, whose units are 1/W$^2$, is reported[2] in dB. We chose PM-QPSK as modulation format because the strength of the non-Gaussianity correction is maximum, since its coefficient $\Phi$ in Eq. 3 is the largest among QAM formats. We show 3-channel systems in Fig. 1 and 15-channel systems in Fig. 2. The reason for choosing these channel numbers is that it was the largest channel number *range* that we could cover through simulations. We also have intermediate sets run at 5 and 9 channels, not shown here both for brevity and because their results are qualitatively very similar to those reported here.

A common feature of all these plots is that the EGN model shows very good accuracy in estimating NLI, throughout all system configurations, confirming the findings in [2] and confirming itself as a reliable reference benchmark.

The GN model always overestimates NLI, along the lines of what was found in [2], [5]-[8]. The extent of the overestimation depends on fiber dispersion and behaves in a peculiar way. The higher the dispersion, the greater the error for low span count, but the lower for high span count. In fact, at 15 channels, SMF is the fiber for which the GN model shows both the highest 1-span error (5.9 dB) and the lowest 50-span error (1.3 dB).

The approximate EGN model Eq. 1, relying on the simple correction Eq. 3, is quite effective with all fibers, showing good convergence towards the exact EGN model curve and vs. simulations, as the number of spans grows. As a result of its asymptotic behavior, Eq. 3 only partially corrects the GN model at low span count. On the other hand, at span counts that are typically of interest for maximum reach predictions, its accuracy is good.

## IV. System performance prediction

The main declared goal of many of the recent modeling efforts has been that of providing a practical tool for realistic system performance prediction. In this section we present a comparison of the accuracy of the GN model and of the approximate EGN model Eq. 1 in predicting maximum system reach in some typical scenarios.

---

[2] All the plots in Figs. 1-2 display the quantity $\eta$, whose units are W$^{-2}$, in dB referred to unity in the specified units, that is, vs. 1·W$^{-2}$.

Note that the EGN model accuracy in predicting system maximum reach was tested in [2], Sect. 6, and found to be excellent, at least in the tested cases, which are the same as those addressed there. Specifically, they are 15-channel PM-QPSK and PM-16QAM systems, running at 32 GBaud. We considered the following channel spacings: 33.6, 35, 40, 45 and 50 GHz. The spectrum was root-raised-cosine with roll-off 0.05. The target BERs were $1.7 \cdot 10^{-3}$ and $2 \cdot 10^{-3}$ respectively, found by assuming a $1 \cdot 10^{-2}$ FEC threshold, decreased by 2 dB of realistic OSNR system margin. EDFA amplification was assumed, with 5 dB noise figure. The considered fibers were: SMF, NZDSF and LS, with the same parameters as before, except for SMF whose loss was set to $\alpha_{dB}=0.2$ dB/km. In addition, we considered pure-silica-core fiber (PSCF) with the following parameters: $D=20.1$ ps/(nm·km), $\gamma=0.8$ (W·km)$^{-1}$, $\alpha_{dB}=0.17$ dB/km.

We point out that we did not assume that the spectrum of NLI was flat, i.e., we did not use the so-called 'white-noise approximation'. We did take into account its actual shape when estimating the system maximum reach, either based on the GN model alone or based on Eq. 1. Note though that, as pointed out in Sect. II, the approximate correction Eq. 3 is assumed frequency-independent. We also point out that the simulative results of this section are found by adding all ASE noise at the end of the link, rather than in-line. The reason for this is that here we want to validate an approximate model that neglects the interaction of in-line ASE noise with non-linearity. Not plotted (for the sake of clarity), the simulative data points with in-line ASE noise are on average about 0.15 dB lower (on $N_{span}$) for PM-QPSK. The effect on PM-16QAM is instead negligible, because PM-16QAM requires a much higher OSNR at the receiver and hence much less ASE noise propagates along the link than for PM-QPSK.

Fig. 3 shows a plot of maximum system reach vs. channel spacing. The GN model underestimates the maximum reach by 0.3-0.6 dB over PSCF, SMF and NZSDF, and up to 0.8 dB over the ultra-low dispersion LS, in agreement with [2], [3], [5]. These results are also in line with the general picture that emerges from Fig. 1-Fig. 2, when taking into account that an error of $x$ dB in the estimation of NLI power leads to an error of about $x/3$ dB in maximum reach estimation [3].

The approximate EGN model 'App. EGN-1' is generated using Eq. 1 in [1], which neglects the non-Gaussianity correction for SCI contribution. With all fibers, for low frequency spacing (33.6 and 35 GHz) the predictions based on it come within a quite small error range $[-0.2, 0]$ dB across all scenarios. The error range widens slightly to $[-0.4, -0.1]$ dB for the larger frequency spacings. This error can be explained by the fact that SCI is overestimated by the App. EGN-1, because that formula does not include SCI non-Gaussianity correction. The fact that it is larger for large frequency spacing confirms this interpretation, because at large frequency spacing SCI represents a greater part of NLI.

With all fibers and spacings, the approximate EGN model

'App. EGN-2' (Eq. 1 of this paper) provides very good accuracy. The error is less than 0.2 dB across all system configurations. We point out that Monte-Carlo simulation uncertainty can be responsible for such very small fluctuations too.

## V. Conclusion

We have presented a compact, closed-form simple correction to the GN model, based on an asymptotic (in the number of spans) approximation of the very accurate but complex EGN model [2].

The formula improves the GN model accuracy by suppressing its tendency to overestimate non-linearity. We have provided quite extensive validation. Albeit approximate, the formula is firmly based on theory and it proves very effective.

In summary, this approximate asymptotic EGN model formula provides a very effective tool that substantially improves the overall accuracy of the GN model in predicting realistic WDM system performance without significantly increasing its computational complexity.

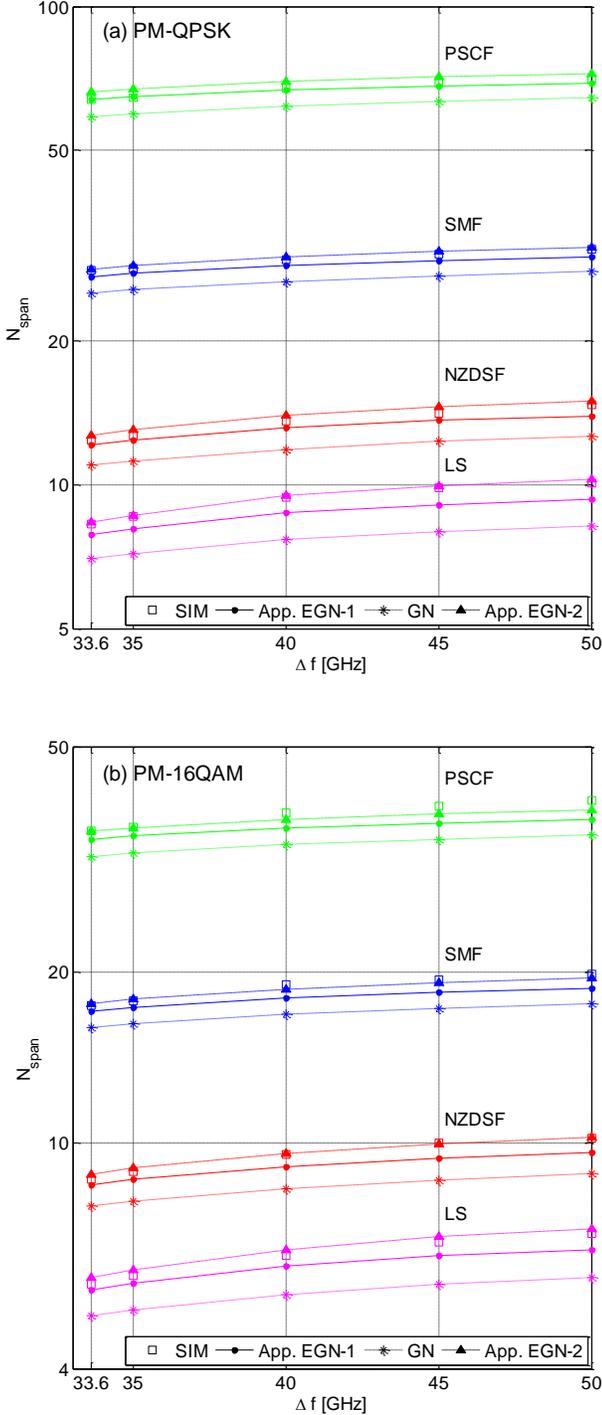

Fig. 3: Plot of maximum system reach for 15-channel PM-QPSK and PM-16QAM systems at 32 GBaud, roll-off 0.05, vs. channel spacing $\Delta f$, over four different fiber types: PSCF, SMF, NZDSF and LS. The span length is 120 km for PM-QPSK and 85 km for PM-16QAM. The 'App. EGN-1' curve is generated using Eq.1 in [1]. The 'App. EGN-2' curve is generated using Eq. 1 in this paper.